\documentclass[journal]{IEEEtran}

\usepackage{url}
\usepackage{array}
\usepackage{cancel}
\usepackage{siunitx}
\usepackage{multirow}
\usepackage{graphicx}
\usepackage{textcomp}
\usepackage{multirow}
\usepackage{stfloats}
\usepackage{algorithm}
\usepackage[T1]{fontenc}
\usepackage[10pt]{moresize}
\usepackage[utf8]{inputenc}
\usepackage[noadjust]{cite}
\usepackage{xcolor,colortbl}
\usepackage{mathtools, cuted}
\usepackage[noend]{algpseudocode}
\usepackage{amsmath,amssymb,amsfonts}

\IEEEoverridecommandlockouts

\graphicspath{{./figs/}}
\newlength{\textfloatsepsave} 
\newcolumntype{L}[1]{>{\raggedright\let\newline\\\arraybackslash\hspace{0pt}}m{#1}}
\newcolumntype{C}[1]{>{\centering\let\newline\\\arraybackslash\hspace{0pt}}m{#1}}
\newcolumntype{R}[1]{>{\raggedleft\let\newline\\\arraybackslash\hspace{0pt}}m{#1}}

\begin{document}
	
	\title{Multiplierless Design of Very Large Constant Multiplications in Cryptography}
	
	\author{Levent Aksoy,~\IEEEmembership{Member,~IEEE,} Debapriya Basu Roy,~\IEEEmembership{Member,~IEEE,} Malik Imran,~\IEEEmembership{Student Member,~IEEE,} Patrick Karl, and~Samuel Pagliarini,~\IEEEmembership{Member,~IEEE}
		
		\thanks{This work has been partially conducted in the project ``ICT programme'' which was supported by the European Union through the European Social Fund. It was also partially supported by the European Union's Horizon 2020 research and innovation programme under grant agreement No 952252 (SAFEST), and by the Estonian Research Council grant MOBERC35.}
		
		\thanks{L.~Aksoy, M.~Imran, and S.~Pagliarini are with the Department of Computer Systems, Centre for Hardware Security, Tallinn University of Technology, Tallinn, Estonia (e-mail: \{levent.aksoy, malik.imran, and samuel.pagliarini\}@taltech.ee.)}
		
		\thanks{D.~B.~Roy and P.~Karl are with the Technical University of Munich, Department of Electrical and Computer Engineering, Chair for Security in Information Technology, Munich, Germany (e-mail: \{debapriya.basu-roy and patrick.karl\}@tum.de.)}
	}
	
	\maketitle
	
	\begin{abstract}
		This brief addresses the problem of implementing very large constant multiplications by a single variable under the shift-adds architecture using a minimum number of adders/subtractors. Due to the intrinsic complexity of the problem, we introduce an approximate algorithm, called {\sc t\~{o}ll}, which partitions the very large constants into smaller ones. To reduce the number of operations, {\sc t\~{o}ll} incorporates \mbox{graph-based} and common subexpression elimination methods proposed for the shift-adds design of constant multiplications. It can also consider the delay of a multiplierless design defined in terms of the maximum number of operations in series, i.e., the number of adder-steps, while reducing the number of operations. \mbox{High-level} experimental results show that the \mbox{adder-steps} of a shift-adds design can be reduced significantly with a little overhead in the number of operations. \mbox{Gate-level} experimental results indicate that while the shift-adds design can lead to a 36.6\% reduction in \mbox{gate-level} area with respect to a design using a multiplier, the delay-aware optimization can yield a 48.3\% reduction in minimum achievable delay of the \mbox{shift-adds} design when compared to the \mbox{area-aware} optimization.
	\end{abstract}
	
	
	\section{Introduction}

Multiplication of constant(s) by a variable is a ubiquitous operation in many applications, such as digital signal processing and cryptography. Since constants are determined beforehand in these applications and the implementation of a multiplier in hardware is expensive in terms of area and power consumption, the constant multiplication can be realized under the shift-adds architecture using only shifts and adders/subtractors~\cite{nguyen}. Note that shifts by a constant value can be realized using only wires which represent no hardware cost. In cryptographic algorithms, such as elliptic curve cryptography (ECC)~\cite{roy19, safecurves} and supersingular isogeny key encapsulation (SIKE)~\cite{roy20, sike}, prime numbers to be multiplied by a variable can respectively be 204-521 bits and 448-768 bits long due to security requirements. The parallel realization of such constant multiplications is required for high-performance cryptographic designs~\cite{roy19}. Thus, the \textit{very large constant multiplication} (VLCM) problem is defined as finding a minimum number of adders/subtractors which realize the multiplication of given very large constants by a variable. Similar to~\cite{np-complete}, this problem is \mbox{NP-complete}.


Techniques under the residue residue number system~\cite{chaves07,low13, patronik17}, that enable large constant multiplications to be realized using a set of small constant multiplications, have been introduced, but they require the logic for conversions between binary and residue number system.  Many large integer multiplication architectures~\cite{karatsuba63, montgomery85, ciara17} have also been proposed, but both operands in these architectures are assumed to be variable. Moreover, prominent algorithms~\cite{jason-dac, spiral, elsevier09, kumm-ilpmcm} have been developed for the shift-adds design of constant multiplications, but they are limited with the bit-width of constants. Furthermore, the VLCM problem has not been studied thoroughly. Hence, we introduce \textbf{the first approximate algorithm {\sc t\~{o}ll} proposed for the VLCM problem}, which is the main contribution of this brief. {\sc t\~{o}ll} divides the very large constants into small coefficients with a reasonable \mbox{bit-width} and re-defines these very large constants as linear equations in the form of summation of shifted versions of these small coefficients. It finds common partial products in a shift-adds design of these small coefficient multiplications using a prominent graph-based (GB) algorithm~\cite{elsevier09,spiral}. It extracts common subexpressions among the linear equations using an efficient common subexpression elimination (CSE) algorithm~\cite{hartley96, hosangadi1}. The performance of a design can be more critical than other characteristics and thus, an increase in area and power consumption can be compromised to meet the performance criterion. Hence, {\sc t\~{o}ll} can also consider the maximum number of operations in series, called the number of \mbox{adder-steps}, while reducing the number of operations. Experimental results show that \mbox{shift-adds} designs obtained by {{\sc t\~{o}ll}} have significantly less hardware complexity than those including generic multipliers and compressor trees, and delay-aware optimization leads to a significant reduction in minimum delay of a design with respect to area-aware optimization.

The remainder of this brief is organized as follows: Section~\ref{sec:background} introduces background concepts. {\sc t\~{o}ll} is described in detail in Section~\ref{sec:algorithm}. Experimental results are given in Section~\ref{sec:results}. Finally, Section~\ref{sec:conclusions} concludes the brief.

	\section{Background}
\label{sec:background}

This section presents background concepts on the \mbox{shift-adds} design of constant multiplications. Since constants are multiplied by a common variable, the realization of constant multiplications corresponds to the realization of constants. For example, $3x = x \ll 1 + x =  (1 \ll 1 + 1)x$ can be rewritten as $3 = 1 \ll 1 +1$ by eliminating the variable $x$ from both sides. These notations will be used interchangeably in this brief.

\begin{figure}[t]
	\centering
	\includegraphics[width=8.0cm]{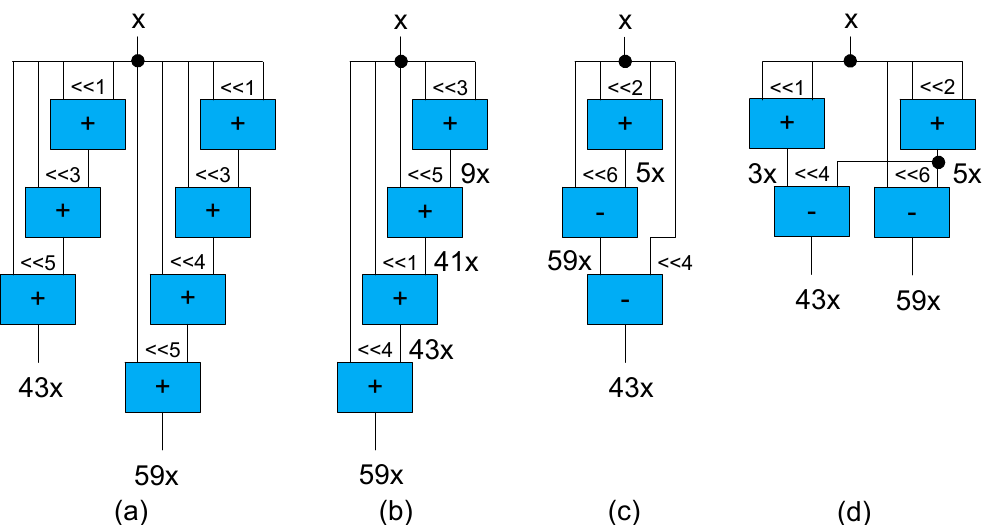}
	\vspace*{-3.4mm}
	\caption{Shift-adds designs of $43x$ and $59x$: (a)~DBR technique~\cite{ercegovac}; (b)~exact CSE method~\cite{tcad08}; (c)~exact GB method~\cite{elsevier09}; (d)~approximate GB method under a delay constraint~\cite{dsd10}.}
	\label{fig:mcm}
	\vspace*{-6mm}
\end{figure}

The straightforward digit-based recoding (DBR) technique~\cite{ercegovac} realizes the shift-adds design of constant multiplications in two steps: (i)~define the constants under a particular number representation, e.g., binary or canonical signed digit (CSD)\footnote{An integer can be written in CSD using $k$ digits as $\sum_{i=0}^{k-1} d_i 2^{i}$, where $d_i \in \{1, 0, -1\}$ with $0 \leq i \leq n-1$. Under CSD, nonzero digits are not adjacent and a minimum number of nonzero digits is used.}~\cite{hartley96}; (ii)~for the nonzero digits in the representation of constants, shift the input variable according to digit positions and add/subtract the shifted variables with respect to digit values. Consider the multiple constant multiplication (MCM) block realizing $43x$ and $59x$ as an example. The decompositions of its constants under binary are given as follows:
\begin{align}
	43x & = (101011)_{bin}x = x\!\ll\!5 + x\!\ll\!3 + x\!\ll\!1 + x \nonumber \\
	59x & = (111011)_{bin}x = x\!\ll\!5 + x\!\ll\!4 + x\!\ll\!3 + x\!\ll\!1 + x \nonumber
\end{align}
which lead to a design with 7 operations in 4 adder-steps, as shown in Fig.~\ref{fig:mcm}(a). 

Algorithms, that aim to maximize the sharing of partial products in the shift-adds design of constant multiplications, can be grouped in two categories based on the search space they explore: (i)~The CSE methods~\cite{tcad08,hartley96,ho08,hosangadi1,lefevre01,park,potkonjak96} initially define the constants under a number representation. Then, in an iterative fashion, after all possible subexpressions that can be extracted from the nonzero digits in representations of constants, are identified, the ``best" subexpression, generally, the most common one, is chosen to be shared among the constant multiplications. The exact CSE algorithm~\cite{tcad08} uses a 0-1 integer linear programming (ILP)-based approach to maximize the sharing of subexpressions. (ii)~The GB methods~\cite{dempster1,dempster4,oscar-diff,jason-dac,spiral,elsevier09,kumm-ilpmcm}, which are not restricted to any particular number representation, aim to find the ``best" intermediate constants, generally, the ones that enable to realize the constant multiplications with a small number of operations. They consider a large number of possible realizations of a constant and obtain better solutions than CSE methods~\cite{elsevier09}. While the exact GB algorithm of~\cite{elsevier09} can explore the search space using \mbox{breadth-first} and depth-first search techniques, the exact GB algorithm of~\cite{kumm-ilpmcm} uses a 0-1 ILP-based approach. 


Returning to our simple MCM example, the exact CSE algorithm~\cite{tcad08} finds a solution with 4 operations in 4 adder-steps when constants are defined under binary, sharing the common subexpressions $9x = (1001)_{bin}x$ and $41x = (101001)_{bin}x$ among the constant multiplications as shown in Fig.~\ref{fig:mcm}(b). On the other hand, the exact GB algorithm~\cite{elsevier09} obtains a solution with a minimum number of 3 operations in 3 adder-steps, finding the intermediate constant multiplication $5x$ to realize the constant multiplications as shown in Fig.~\ref{fig:mcm}(c). 

In a shift-adds design of constant multiplications, the delay is generally defined as the number of adder-steps~\cite{kang}. Note that the minimum adder-steps of a single constant $c$ is computed as $mas_c = \lceil log_2 NZ(c)\rceil$, where $NZ(c)$ denotes the number of nonzero digits in the CSD representation of the constant. Thus, given a set of constants $C = \{c_1, c_2, \ldots, c_n\}$, the minimum adder-steps of multiple constants in the set $C$ is computed as $mas_C =\max_{1 \leq i \leq n}\{mas_{c_i}\}$~\cite{kang}. There exist efficient CSE and GB algorithms introduced to optimize the number of operations in the multiplierless design where the delay constraint given in terms of the number of adder-steps is never violated~\cite{tcad08, kang, dempster5, dsd10}. Returning to our example, Fig.~\ref{fig:mcm}(d) shows the realization of constant multiplications with a minimum number of adder-steps, i.e., 2, whose solution is obtained by the GB algorithm of~\cite{dsd10} using 4 operations.

The proposed algorithms, except the DBR technique, are limited with the size of constants. This is simply because the number of possible partial products of a constant increases dramatically as its bit-width increases~\cite{elsevier09}. For example, the exact GB algorithm~\cite{jason-dac} developed for the shift-adds design of a single constant multiplication can handle a constant up to 32 bits. The approximate~\cite{spiral} and exact~\cite{elsevier09} GB algorithms can handle multiple constants up to 31 and 16 bits, respectively. 


	\section{{\sc t\~{o}ll} - The Proposed Method}
\label{sec:algorithm}

{\sc t\~{o}ll} takes $n$ large constants, i.e., $lc_1, lc_2, \ldots, lc_{n}$, in hexadecimal format and the number of bits in partition, i.e., $p$, as inputs and returns the multiplication of these large constants by an input variable under the \mbox{shift-adds} architecture described in Verilog as an output. Due to the limitations of algorithms proposed for the multiplierless design on the size of constants, the value of $p$ is determined to be a multiple of 4 with a minimum and maximum value of 4 and 28, respectively. It initially partitions the large constants into $p$-bit coefficients\footnote{Partitioning of a $k$-bit large constant $lc$ into $p$-bit coefficients can be written as $\sum_{i=1}^{\lceil k/p \rceil} lc[ip-1 : (i-1)p] 2^{(i-1)p} = \sum_{i=1}^{\lceil k/p \rceil} c_i 2^{(i-1)p}$.} and defines each large constant as the summation of shifted \mbox{$p$-bit} coefficients, called a linear equation. Then, it applies a GB algorithm~\cite{spiral, elsevier09} to these coefficients to find their multiplierless realization. Finally, it uses a CSE heuristic~\cite{hartley96, hosangadi1} to extract common subexpressions in the linear equations and realizes the final linear equations using \mbox{two-term} subexpressions. It includes three stages: (i)~partitioning; (ii)~realization of coefficients; and (iii)~realization of linear equations. It can also consider the delay of the multiplierless design while reducing the number of operations. In following, its stages are described under the area-aware optimization. Finally, details in the delay-aware optimization are given.

\begin{figure}[t]
	\centering
	\vspace*{-0mm}
	\includegraphics[width=8.8cm]{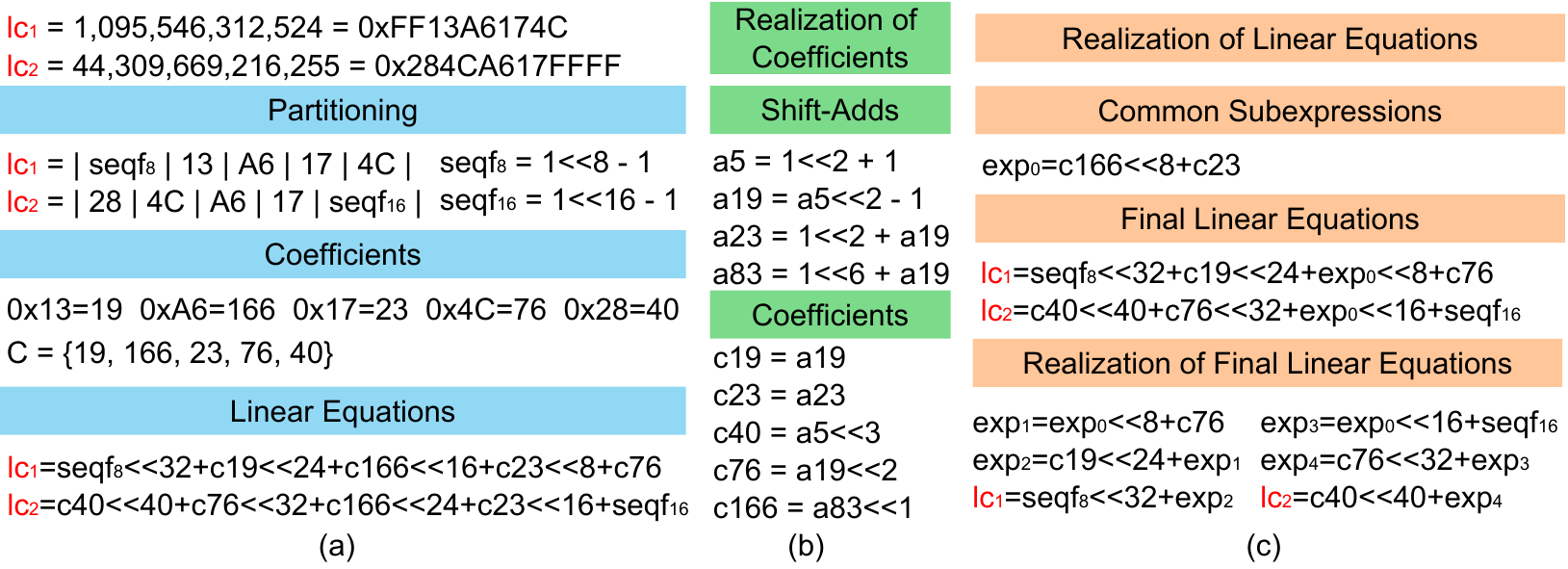}
	\vspace*{-8mm}
	\caption{Stages of {\sc t\~{o}ll} on a small example with two large constants: (a)~partitioning; (b)~realization of coefficients; (c)~realization of linear equations.}
	\label{fig:toll}
	\vspace*{-5mm}
\end{figure}

\textbf{Partitioning:} In {\sc t\~{o}ll}, two partitioning strategies are implemented. In the first one, called the \textit{strict} partitioning, starting from the least significant bit, $p$-bit coefficients are generated from the hexadecimal digits of each large constant and stored as integers in set $C$ without repetition. Shift values of these coefficients are computed based on the locations of hexadecimal digits and stored in set $S$. While partitioning large constants into \mbox{$p$-bit} coefficients, sequences of $r$ 0s, where $r \geq p$ and \mbox{$r \mod{p} = 0$}, are found and ignored, since such a sequence requires no operations. Also, sequences of $r$ 1s, where $r \geq p$  and \mbox{$r \mod{p} = 0$}, are identified and replaced by a subexpression, denoted as $seqf_r$, which needs only a single subtractor, i.e., $2^{r}-1$. These sequence subexpressions are stored in set $Seqf$ without repetition. Finally, the realization of each large constant is written as a linear equation in the form of summation of coefficients in set $C$ based on their shift values in set $S$ and sequence expressions in set $Seqf$ based on their shift values in large constants. Fig.~\ref{fig:toll}(a) shows the steps of the \textit{strict} partitioning strategy when $p$ is 8. 

\begin{figure}[t]
	\centering
	\includegraphics[width=8.0cm]{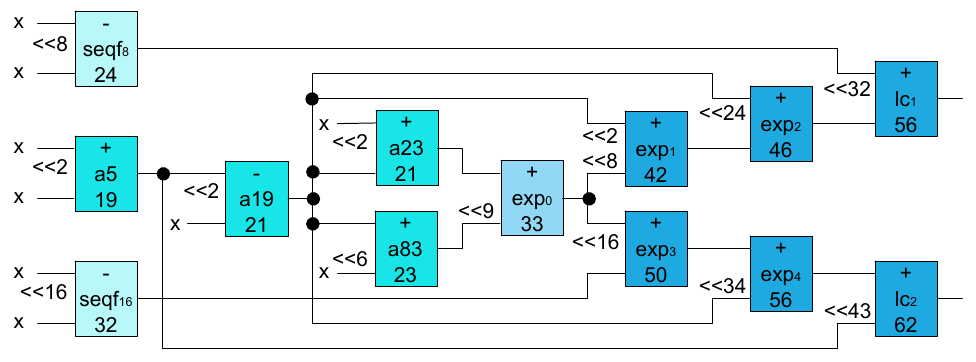}
	\vspace*{-4mm}
	\caption{Multiplierless realization of large constant multiplications in Fig.~\ref{fig:toll}.}
	\label{fig:rea}
	\vspace*{-7mm}
\end{figure}

In the second one, called \textit{common digit} partitioning, initially, all possible $p$-bit coefficients are identified and the ones, which occur more than once, are extracted from the large constant in an order of their number of occurrences iteratively, starting from the greatest one. Then, the remaining digits are divided based on the \textit{strict} partitioning. The \textit{common digit} partitioning aims to increase the sharing of common expressions while realizing the linear equations. For our example, common coefficients $0\!\times\!A6$, $0\!\times\!17$, and $0\!\times\!4C$ are initially extracted.

\textbf{Realization of Coefficients:} Coefficients in the linear equations of each large constant determined at the partitioning stage and stored in set $C$ are realized under the shift-adds architecture. {\sc t\~{o}ll} incorporates two prominent GB algorithms~\cite{spiral, elsevier09}. For our example, Fig.~\ref{fig:toll}(b) presents the solution of the exact algorithm~\cite{elsevier09} on coefficients in set $C$ with 4 operations in 3 adder-steps.

\textbf{Realization of Linear Equations:} In {\sc t\~{o}ll}, common subexpressions in the linear equations obtained at the partitioning stage are identified and eliminated using a CSE heuristic. The developed CSE method is based on the CSE heuristics of~\cite{hartley96, hosangadi1}. In this method, all subexpressions with two terms are found considering their shift values, their number of occurrences is computed, and the one with the maximum number of occurrences greater than 1 is chosen to be eliminated. This process is iterated until there is no subexpression with a maximum number of occurrences greater than 1. For our example, Fig.~\ref{fig:toll}(c) shows the realization of the common subexpression and the final version of linear equations after this subexpresion is eliminated. Then, till the number of coefficients and subexpressions in each final linear equation is less than or equal to 2, in an iterative fashion, two terms having the smallest bit-width value are determined, defined as a subexpression indicating the summation of these terms, and eliminated from the linear equations. The insight behind the selection of coefficients and subexpressions based on their bit-widths is to reduce design area. For our example, Fig.~\ref{fig:toll}(c) also shows the realization of final linear equations which needs 6 operations in 3 adder-steps. 

Observe from Fig.~\ref{fig:toll} that the multiplierless design requires a total number of 13 adders/subtractors, i.e., 2 for the sequence subexpressions, 4 for the shift-adds realization of coefficients, 1 for the common subexpression, and 6 for the realization of final linear equations, in 7 adder-steps. Fig.~\ref{fig:rea} presents this multiplierless design, where the \textit{top}, \textit{middle}, and \textit{bottom} terms inside operations stand respectively for the operation type, operation output, and bit-width of the operation output, assuming that the input variable $x$ is 16 bits long.

\begin{table}[t]
	\centering
	\scriptsize
	\caption{High-level results of shift-adds designs realizing single prime number multiplications.}
	\vspace*{-3mm}
	\begin{tabular}{|@{\hskip3pt}L{18mm}@{\hskip3pt}|@{\hskip3pt}C{6.1mm}@{\hskip3pt}|C{4.4mm}@{\hskip3pt}|@{\hskip3pt}C{4mm}@{\hskip3pt}|@{\hskip3pt}c@{\hskip3pt}|@{\hskip3pt}C{4.4mm}@{\hskip3pt}|@{\hskip3pt}C{4mm}@{\hskip3pt}|@{\hskip3pt}c@{\hskip3pt}|@{\hskip3pt}C{4.4mm}@{\hskip3pt}|@{\hskip3pt}C{4mm}@{\hskip3pt}|@{\hskip3pt}c@{\hskip3pt}|}
		\hline
		\multirow{2}{*}{Instance} & Prime & \multicolumn{3}{c|@{\hskip3pt}}{$p=8$} & \multicolumn{3}{c|@{\hskip3pt}}{$p=16$} & \multicolumn{3}{c|}{$p=24$}\\
		\cline{3-11}
		& Width  & oper & step & time & oper & step & time & oper & step & time \\
		\hline \hline
		anomalous~\cite{safecurves}      & 220 & 19 & 13 & 4.5  & 15 & 12 & 4.5  & 15 & 10 & 4.8  \\
		anssifrp~\cite{safecurves}       & 272 & 60 & 33 & 5.7  & 43 & 24 & 5.6  & 43 & 15 & 6.8  \\
		bn(2,254)~\cite{safecurves}      & 268 & 39 & 22 & 5.5  & 32 & 16 & 5.4  & 29 & 18 & 5.7  \\
		brainpool256~\cite{safecurves}   & 268 & 58 & 35 & 5.6  & 44 & 27 & 5.5  & 41 & 20 & 7.3  \\
		brainpool348~\cite{safecurves}   & 400 & 80 & 51 & 8.1  & 61 & 33 & 7.9  & 54 & 27 & 10.4 \\
		sike610~\cite{sike}              & 640 & 69 & 39 & 12.4 & 51 & 28 & 12.5 & 47 & 24 & 14.3 \\
		sike751~\cite{sike}              & 768 & 87 & 50 & 14.7 & 62 & 36 & 14.7 & 55 & 27 & 17.3 \\
		\hline
	\end{tabular}
	\label{tab:high-primes}
	\vspace*{-8mm}
\end{table}

\textbf{Delay-Aware Optimization:} During the delay-aware realization of large constants under the shift-adds architecture, the coefficients of linear equations determined at the partitioning stage are realized using the algorithms of~\cite{spiral, dsd10} when the delay constraint is set to the $mas_C$ value of multiple coefficients in set $C$. For our example, the coefficients are implemented using 5 operations in 2 \mbox{adder-steps} using the algorithm of~\cite{dsd10}. Moreover, while choosing common subexpressions among the ones with the maximum number of occurrences to be eliminated in the linear equations, the one, that leads to the smallest increase in the number of adder-steps, is preferred. For our example, the subexpression $exp_0$ of Fig.~\ref{fig:toll} is also selected during the \mbox{delay-aware} optimization. Lastly, in the realization of final linear equations, the subexpressions are generated with a minimum number of adder-steps considering the \mbox{bit-width} of coefficients and subexpressions. For our example, the final linear equations are realized using 6 operations in 2 \mbox{adder-steps}. As an example, the final linear equation $lc_1$ is realized as $lc_1 = exp_2 + exp_1$, where $exp_1 = exp_0<<8 + c76$ and $exp_2 = seqf_{8}<<32 + c19<<24$. We note that for our example, the delay-aware optimization leads to a design with a total number of 14 operations in 5 adder-steps. 

In {\sc t\~{o}ll}, the design and verification process of the multiplierless realization of large constant multiplications is automated. {\sc t\~{o}ll} can generate the behavioral description of the design in Verilog and the associated testbench for verification. It can also describe the large constant multiplications using multipliers in Verilog. {\sc t\~{o}ll} is available at \textit{https://github.com/leventaksoy/vlcm}.
	\section{Experimental Results}
\label{sec:results}

\begin{table*}[t]
	\centering
	\scriptsize
	\caption{Gate-level results of designs realizing single prime number multiplications.}
	\vspace*{-3mm}
	\begin{tabular}{|l|c|c|c|c|c|c|c|c|c|c|c|c|c|c|c|}
		\hline
		\multirow{3}{*}{Instance} & \multicolumn{3}{c|}{Generic} & \multicolumn{3}{c|}{Compressor} & \multicolumn{9}{c|}{Shift-Adds} \\
		\cline{8-16}
		& \multicolumn{3}{c|}{Multiplier} & \multicolumn{3}{c|}{Trees} & \multicolumn{3}{c|}{$p=8$} & \multicolumn{3}{c|}{$p=16$} & \multicolumn{3}{c|}{$p=24$}\\
		\cline{2-16}
		& area & delay & power & area & delay & power & area & delay & power & area & delay & power & area & delay & power \\
		\hline \hline
		anomalous~\cite{safecurves}      & 3477  & 9846  & 849  & 4367  & 4742  & 161 & 2516  & 4234  & 677  & 2202  & 4480  & 606  & 2530  & 5922  & 823  \\
		anssifrp~\cite{safecurves}       & 9525  & 23215 & 2630 & 12339 & 27401 & 526 & 8082  & 31183 & 2441 & 7988  & 25890 & 2524 & 9506  & 24351 & 2998 \\
		bn(2,254)~\cite{safecurves}      & 6297  & 22150 & 1537 & 8585  & 11266 & 322 & 4929  & 9460  & 1335 & 4968  & 10646 & 1505 & 4957  & 10381 & 1474 \\
		brainpool256~\cite{safecurves}   & 10292 & 22956 & 2815 & 12356 & 27615 & 540 & 8268  & 32319 & 2583 & 8273  & 27506 & 2738 & 8779  & 23264 & 2906 \\
		brainpool348~\cite{safecurves}   & 14539 & 33738 & 4010 & 17184 & 37916 & 759 & 10942 & 43181 & 3367 & 11095 & 36695 & 3608 & 11280 & 31180 & 3777 \\
		sike610~\cite{sike}              & 11597 & 32475 & 3117 & 14217 & 32933 & 609 & 10476 & 33891 & 3163 & 10751 & 29473 & 3652 & 11562 & 31862 & 4100 \\
		sike751~\cite{sike}              & 14670 & 40180 & 4037 & 17636 & 38633 & 786 & 12757 & 40163 & 3974 & 12649 & 35587 & 4137 & 14453 & 39910 & 5255 \\
		\hline
	\end{tabular}
	\label{tab:gate-primes}
	\vspace*{-7mm}
\end{table*}

\begin{table}[t]
	\centering
	\scriptsize
	\caption{Gate-level results of designs with minimum achievable delay.}
	\vspace*{-3mm}
	\begin{tabular}{|@{\hskip3pt}l@{\hskip2pt}|c|c|@{\hskip3pt}c@{\hskip3pt}|@{\hskip3pt}c@{\hskip3pt}|@{\hskip3pt}c@{\hskip3pt}|@{\hskip3pt}c@{\hskip3pt}|@{\hskip3pt}c@{\hskip3pt}|@{\hskip3pt}c@{\hskip3pt}|}
		\hline
		\multirow{2}{*}{Instance} & \multicolumn{3}{c|@{\hskip3pt}}{Area-Aware Optimization} & \multicolumn{5}{c|}{Delay-Aware Optimization} \\
		\cline{2-9}
		& area & delay & power & oper & step & area & delay & power \\
		\hline \hline
		anomalous~\cite{safecurves}    & 6296  & 1135 & 1304 & 17 & 6 & 5952  & 1116 & 1317 \\
		anssifrp~\cite{safecurves}     & 22488 & 1982 & 5786 & 52 & 8 & 22696 & 1403 & 5152 \\
		bn(2,254)~\cite{safecurves}    & 15069 & 1582 & 3868 & 34 & 7 & 12597 & 1258 & 2885 \\
		brainpool256~\cite{safecurves} & 22937 & 2396 & 6487 & 49 & 8 & 21385 & 1404 & 5312 \\
		brainpool348~\cite{safecurves} & 29394 & 2729 & 7889 & 69 & 8 & 30440 & 1410 & 6714 \\
		sike610~\cite{sike}            & 27444 & 2534 & 8324 & 64 & 8 & 29243 & 1528 & 7601 \\
		sike751~\cite{sike}            & 29736 & 3643 & 8355 & 71 & 8 & 33159 & 2167 & 8899 \\
		\hline
	\end{tabular}
	\label{tab:gate-primes-mad}
	\vspace*{-7mm}
\end{table}

\begin{figure*}[b]
	\centering
	\vspace*{-8mm}
	\parbox{5.8cm}{\centerline{\includegraphics[width=6.5cm]{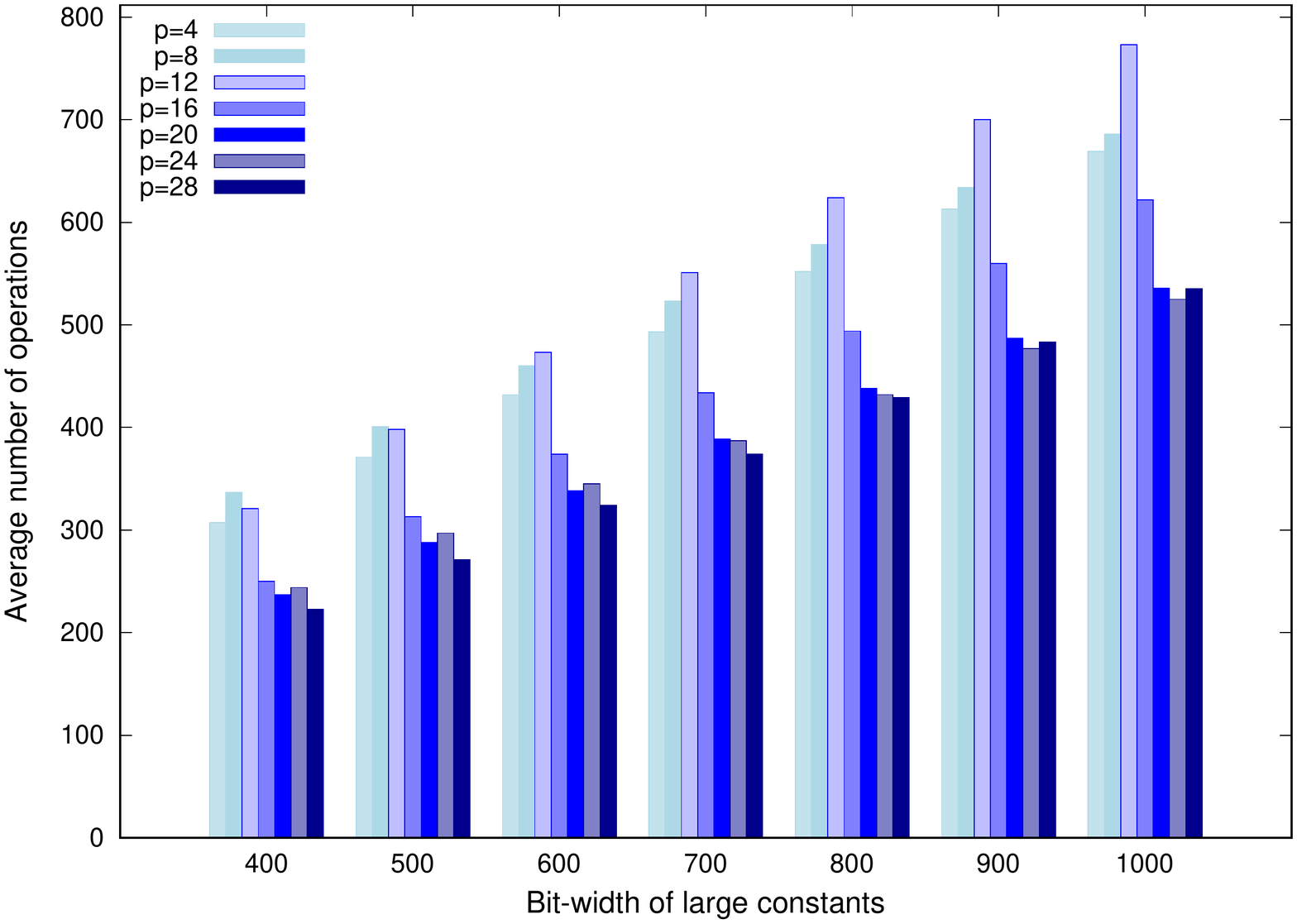}}}\
	\parbox{5.8cm}{\centerline{\includegraphics[width=6.5cm]{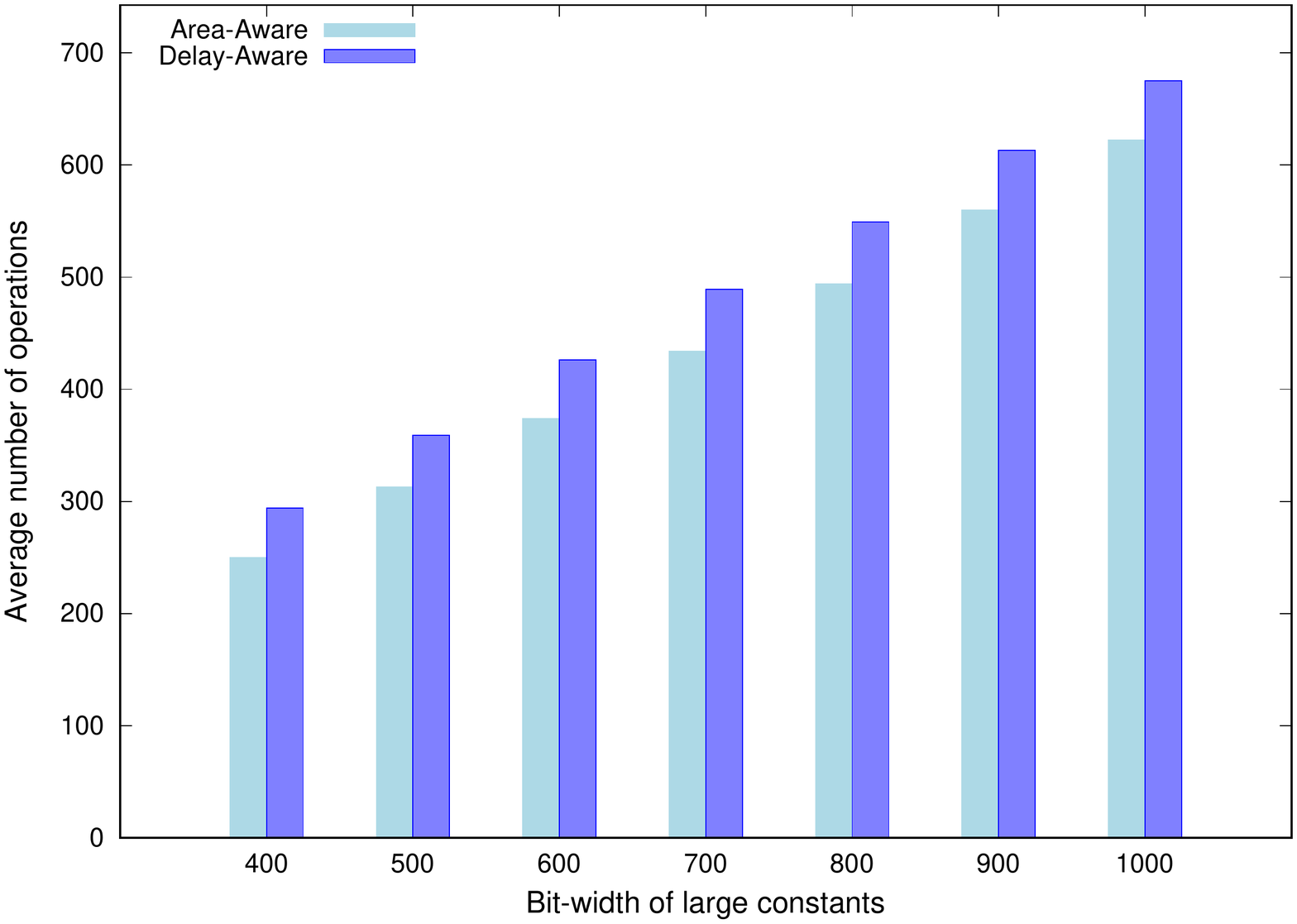}}}\
	\parbox{5.8cm}{\centerline{\includegraphics[width=6.5cm]{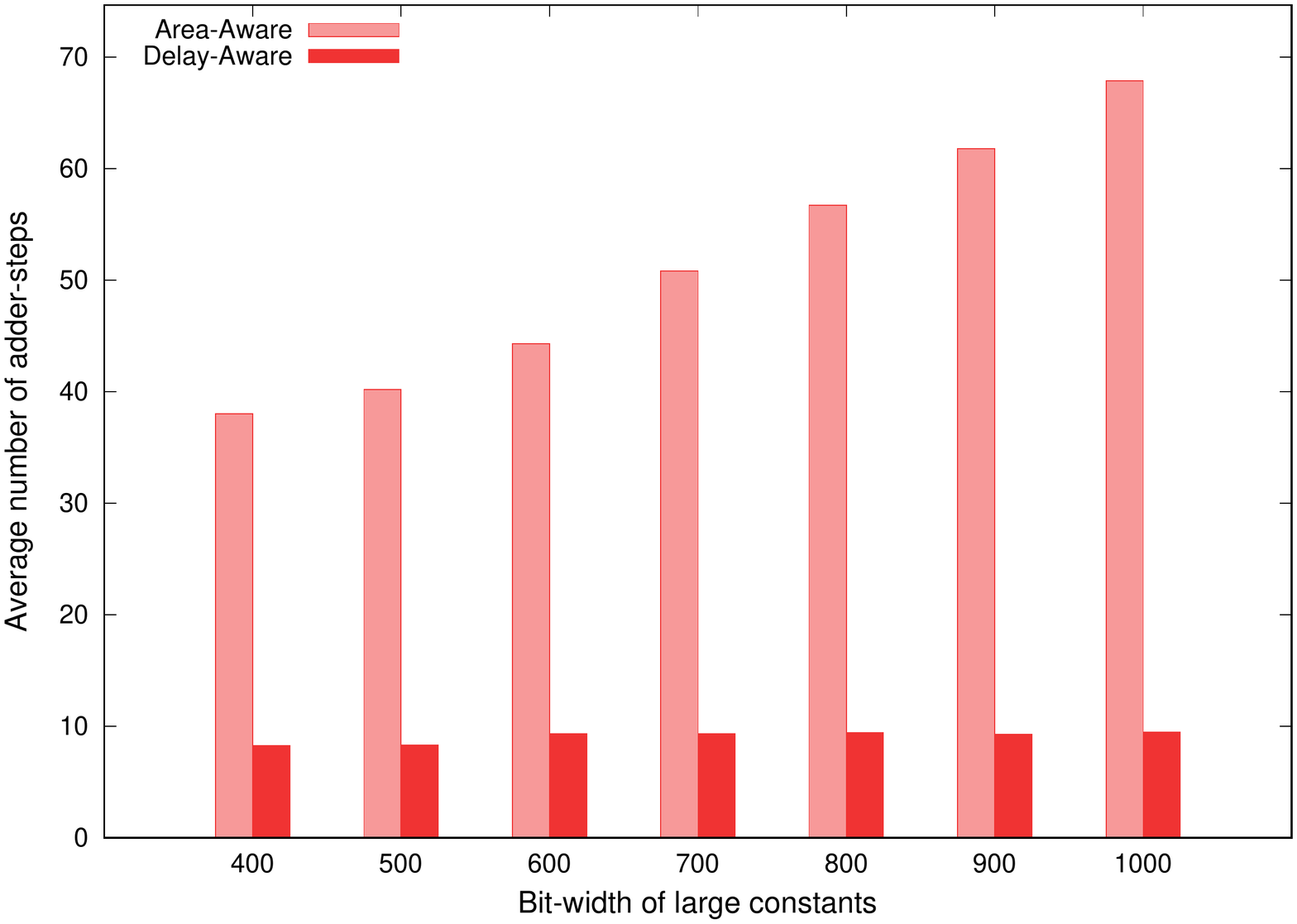}}}\
	
	\vspace*{-3.5mm}
	
	\parbox{5.8cm}{\centerline{\scriptsize (a)}}\
	\parbox{5.8cm}{\centerline{\scriptsize (b)}}\
	\parbox{5.8cm}{\centerline{\scriptsize (c)}}\
	\vspace*{-3mm}
	\caption{Results on randomly generated instances when $n$ is 5: (a)~impact of $p$ on the number of operations; (b)-(c)~impact of the optimization technique on the number of operations and adder-steps when $p$ is 16.}  
	\label{fig:rand-oper}
\end{figure*}

As the first experiment set, the well-known cryptographic prime numbers are taken from~\cite{safecurves,sike} and the related constants to be multiplied by a variable are computed. Table~\ref{tab:high-primes} presents these instances, each of which includes a single prime number, i.e., $n$ is 1, and their bit-width values. Note that these elliptic curves are chosen because the underlying primes do not have any special form. Other elliptic curves, such as \textit{Curve25519} and \textit{NIST Curves}, are based on either pseudo-Mersenne or Solinas primes where modular reductions are performed using a small number of adders/subtractors~\cite{safecurves}. Hence, modular reduction in elliptic curves given in Table~\ref{tab:high-primes} are performed using Montgomery reduction that involves constant multiplication. The results shown in this work focus only on the constant multiplication of Montgomery reduction, not the entire Montgomery reduction. Table~\ref{tab:high-primes} also shows the \mbox{high-level} results of \mbox{shift-adds} designs, where \textit{oper}, \textit{step}, and \textit{time} denote the number of operations, the number of \mbox{adder-steps}, and the run-time of {\sc t\~{o}ll} in seconds, respectively. These results were obtained when the \textit{strict} partitioning strategy is chosen, the area-aware optimization is used, the approximate GB algorithm~\cite{spiral} is selected for the \mbox{shift-adds} realization of coefficients, and the bit-width of the input variable is 16. Note that {\sc t\~{o}ll} was run on a PC including an Intel Core \mbox{i5-10600K} processing unit at 4.1\,GHz with 16\,GB memory. 

Observe from Table~\ref{tab:high-primes} that the use of a high $p$ value leads to a shift-adds design with a small number of operations and the multiplierless realizations are obtained in a reasonable time.

Table~\ref{tab:gate-primes} presents the gate-level results of designs realizing prime number multiplications using a generic multiplier, compressor trees, and adders/subtractors under the shift-adds architecture. In this table, \textit{area}, \textit{delay}, and \textit{power} stand for the total area in $\mu m^2$, delay in the critical path in $ps$, and total power dissipation in $\mu W$, respectively. Logic synthesis was performed by Cadence Genus using a commercial 65\,nm cell library without a strict delay constraint aiming for area optimization. Designs are validated using 10,000 randomly generated inputs in simulation.

Observe from Table~\ref{tab:gate-primes} that the shift-adds designs occupy less area when compared to those using a generic multiplier and compressor trees. The gain in area on the shift-adds design with respect to the one using a generic multiplier (compressor trees) reaches up to 36.6\% (49.5\%) on the \textit{anomalous} instance when $p$ is 16. Note also that as $p$ increases, the hardware complexity of the shift-adds design tends to increase, although there are designs obtained when $p$ is 16 (or 24) with less area when compared to those obtained when $p$ is 8 (or 16). This is because as $p$ increases, the sizes of operations, which have an impact on the hardware complexity, are increased. 

In order to show the impact of the optimization techniques on the minimum achievable delay, the shift-adds designs obtained under the area- and delay-aware optimization when $p$ is 16 are synthesized with timing constraints changed in a binary search manner till the minimum delay in the critical path is found without a negative slack. The initial lower and upper bounds of the timing constraint are taken as 0 and 80 $ns$, respectively. Table~\ref{tab:gate-primes-mad} shows the gate-level results of designs with the minimum achievable delay and also, the high-level results of designs when the delay-aware optimization is used. 

Observe from Tables~\ref{tab:high-primes} and~\ref{tab:gate-primes-mad} that the delay-aware optimization yields significant reduction in the number of \mbox{adder-steps} with an increase in the number of operations. Thus, the designs obtained using the delay-aware optimization have significantly improved delay over those obtained using the area-aware optimization, reaching up to a 48.3\% reduction on the \textit{brainpool348} instance. However, those designs have larger area than the ones obtained under the area-aware optimization, e.g., the \textit{anssifrp} and \textit{brainpool348} instances. It is also observed from the results obtained during the binary search of minimum delay that the delay-aware optimization can generate designs with less area and delay when compared to the design obtained under the area-aware optimization with the minimum delay, e.g., the \textit{bn(2,254)}, and \textit{brainpool256} instances.

As the second experiment set, we used randomly generated multiple constants whose bit-width ranges from 400 to 1000 in a step of 100, i.e., a total of 7 categories. We generated 30 instances for the same bit-width of constants when the number of constants, i.e., $n$, is 5, a total of $7 \times 30 = 210$ instances. In this experiment, the algorithm of~\cite{spiral} is used to realize the multiplierless design of coefficients and the \textit{strict} partitioning strategy is chosen. 

Fig.~\ref{fig:rand-oper}(a) presents the average number of operations obtained by {\sc t\~{o}ll} when the area-aware optimization is considered. Observe that the use of a high $p$ value for partitioning the hexadecimal digits decreases the required number of operations, simply because it reduces the number of terms in linear equations. Interestingly, less number of operations can be obtained when $p$ is decreased, because the number of common subexpressions in the linear equations is increased in this case. Although it is clear that an increase in $p$ decreases the required number of operations, the prominent GB algorithms~\cite{spiral, elsevier09} are limited with the size of coefficients and the reduction of the number of operations does not always lead to a design with a small area as shown in Tables~\ref{tab:high-primes} and~\ref{tab:gate-primes}.

Figs.~\ref{fig:rand-oper}(b)-(c) show the average number of operations and \mbox{adder-steps} obtained by {\sc t\~{o}ll} when the area- and delay-aware optimizations are used and $p$ is 16. Note that the \mbox{delay-aware} optimization can reduce the number of \mbox{adder-steps} of a \mbox{shift-adds} design significantly, but with an increase in the number of operations. Note that while the maximum increase in the number of operations is $1.17\times$, the maximum decrease in the number of adder-steps is $7.14\times$ in the delay-aware optimization with respect to the area-aware optimization.

	\section{Conclusions}
\label{sec:conclusions}

This brief introduced {\sc t\~{o}ll}, the first approximate algorithm proposed for the VLCM problem. Our method is equipped with both area and delay optimization techniques, including previously proposed algorithms used to reduce the number of operations and adder-steps of a \mbox{shift-adds} design. Experimental results clearly indicated that {\sc t\~{o}ll} can lead to a significant reduction on the circuit area when compared to that of a design with a multiplier or compressor trees. It can also generate alternative designs which may help a designer to choose the best fit for the design requirements in a given application.

	\bibliography{lcm}
	\bibliographystyle{IEEEtran}
	
\end{document}